\documentclass[aps,prl,twocolumn,showpacs,superscriptaddress,groupedaddress]{revtex4}
\usepackage{graphicx,dcolumn,bm,amssymb} 

\newcommand{\zco}{ZnCr$_2$O$_4$}
\newcommand{\mco}{MgCr$_2$O$_4$}
\newcommand{\nco}{NiCr$_2$O$_4$}
\newcommand{\cuco}{CuCr$_2$O$_4$}
\newcommand{\cco}{CdCr$_2$O$_4$}
\newcommand{\aco}{$A$Cr$_2$O$_4$}
\newcommand{\hco}{HgCr$_2$O$_4$}

\begin{document}

\title{Crystal structures of spin-Jahn-Teller--ordered \mco\, and \zco\,}

\author{Moureen\,C.\,Kemei}\email{kemei@mrl.ucsb.edu}
\author{Phillip\,T.\,Barton}
\author{Stephanie\,L.\,Moffitt}
\author{Michael\,W.\,Gaultois}
\author{Joshua\,A.\,Kurzman}
\author{Ram\,Seshadri}
\affiliation{Materials Department and Materials Research Laboratory\\
University of California, Santa Barbara, CA, 93106, USA}
\author{Matthew\,R.\,Suchomel}
\affiliation{X-Ray Science Division and Material Science Division\\ 
Argonne National Laboratory, Argonne IL, 60439, USA}
\author{Young-Il Kim}
\affiliation{Department of Chemistry, Yeungnam University\\ 
Gyeongsan, Gyeongbuk 712-749, Korea}

\date{\today}  

\begin{abstract}
Magnetic ordering in the geometrically frustrated magnetic oxide spinels 
\mco\/ and \zco\/ is accompanied by a structural change that helps relieve
the frustration. Analysis of high-resolution synchrotron 
X-ray scattering reveals that the low-temperature structures are well described 
by a two-phase model of tetragonal $I4_1/amd$ and orthorhombic 
$Fddd$ symmetries. The Cr$_4$ tetrahedra of the pyrochlore lattice 
 are distorted at these low-temperatures, with the $Fddd$ phase 
displaying larger distortions than the $I4_1/amd$ phase. The spin-Jahn-Teller
distortion is approximately one order of magnitude smaller than is observed
in first-order Jahn-Teller spinels such as NiCr$_2$O$_4$ and CuCr$_2$O$_4$. 
In analogy with NiCr$_2$O$_4$ and CuCr$_2$O$_4$, we further suggest that the 
precise nature of magnetic ordering can itself provide a second driving force 
for structural change. 

\end{abstract}

\maketitle

The \aco\/ spinels possess highly degenerate spin liquid states that can
order at low temperature in conjunction with a lattice distortion, in a 
manner sometimes referred to as spin-Jahn-Teller ordering.\cite{lee_2000,tchernyshyov_2002}  
Despite extensive studies of the spin-Jahn-Teller phases of \aco\, spinels, 
there is little agreement on the full description of the low-temperature 
structures of \mco\, and \zco.\cite{lee_2007,kagomiya_2002}
At room temperature, \aco\, are cubic spinels in the space group $Fd\overline{3}m$, 
provided the $A$ ions are non-magnetic. $A$ cations occupy tetrahedral sites while Cr$^{3+}$
with spin $S=3/2$ populate octahedral sites. These are normal spinels: Cr$^{3+}$
shows a strong preference for the octahedral site.\cite{dunitz_1957}
Magnetic frustration in \aco\, spinels is known to decrease from $A$ = Zn to Mg to
Cd to Hg with the respective spinels showing Weiss intercepts $\Theta_{CW}$ of
$-$390\,K,\cite{rudolf_2007} $-$346\,K,\cite{rudolf_2007,ramirez_1994} $-$71\,K,\cite{rudolf_2007}
 and $-$32\,K\cite{rudolf_2007,ueda_2006} and spin-Jahn-Teller ordering temperatures
($T_N$) of $\approx$12.7\,K,\cite{martin_2008} $\approx$12.5\,K,\cite{lee_2000}
 $\approx$7.8\,K,\cite{rovers_2002,chung_2005,aguilar_2008} and $\approx$5.8\,K\cite{ueda_2006}. 

Several low-temperature nuclear structures have been proposed for \aco\,
spinels. X-ray diffraction studies reveal $Fddd$ symmetry in the spin-Jahn-Teller 
phase of \hco.\cite{ueda_2006} A tetragonal $I4_1/amd$ structure of \mco\, 
was identified in low-temperature synchrotron X-ray\cite{ehrenberg_2002} and neutron powder 
diffraction studies.\cite{martin_2008}  A tetragonal distortion has also been
observed in the antiferromagnetic phase of \cco, identified by Aguilar
\textit{et al.}\ using infrared spectroscopy,\cite{aguilar_2008} and by Chung
and co-workers from elastic and inelastic neutron scattering
studies.\cite{chung_2005}  The low-temperature structure of \cco\, was 
assigned to the $I4_1/amd$ space group as reported by Lee \textit{et al.} from
synchrotron X-ray and neutron scattering studies of single
crystals.\cite{lee_2007} In the same report, single crystals of ZnCr$_2$O$_4$
were reported to adopt the tetragonal $I\bar4m2$ space group below the N\'eel
temperature.\cite{lee_2007} However, X-ray powder diffraction by Kagomiya
\textit{et al.} suggested, without describing the complete structure, 
that at low-temperatures ZnCr$_2$O$_4$ is
modelled by the orthorhombic space group $Fddd$.\cite{kagomiya_2002} Recent 
electron-spin resonance studies of 
single crystal \zco\, by Glazkov \textit{et\,al.}
showed that tetragonal and orthorhombic distortions coexist in the N\'eel phase
of \zco.\cite{glazkov_2008}  
In this Letter, we report coexisting tetragonal $I4_1/amd$ and orthorhombic
$Fddd$ symmetries in the spin-Jahn-Teller phases of \mco\, and \zco\,,
observed using high-resolution synchrotron X-ray powder
diffraction. Phase coexistence is suggested for the first time in these materials 
from diffraction studies.

\begin{figure}
\includegraphics[scale=0.8]{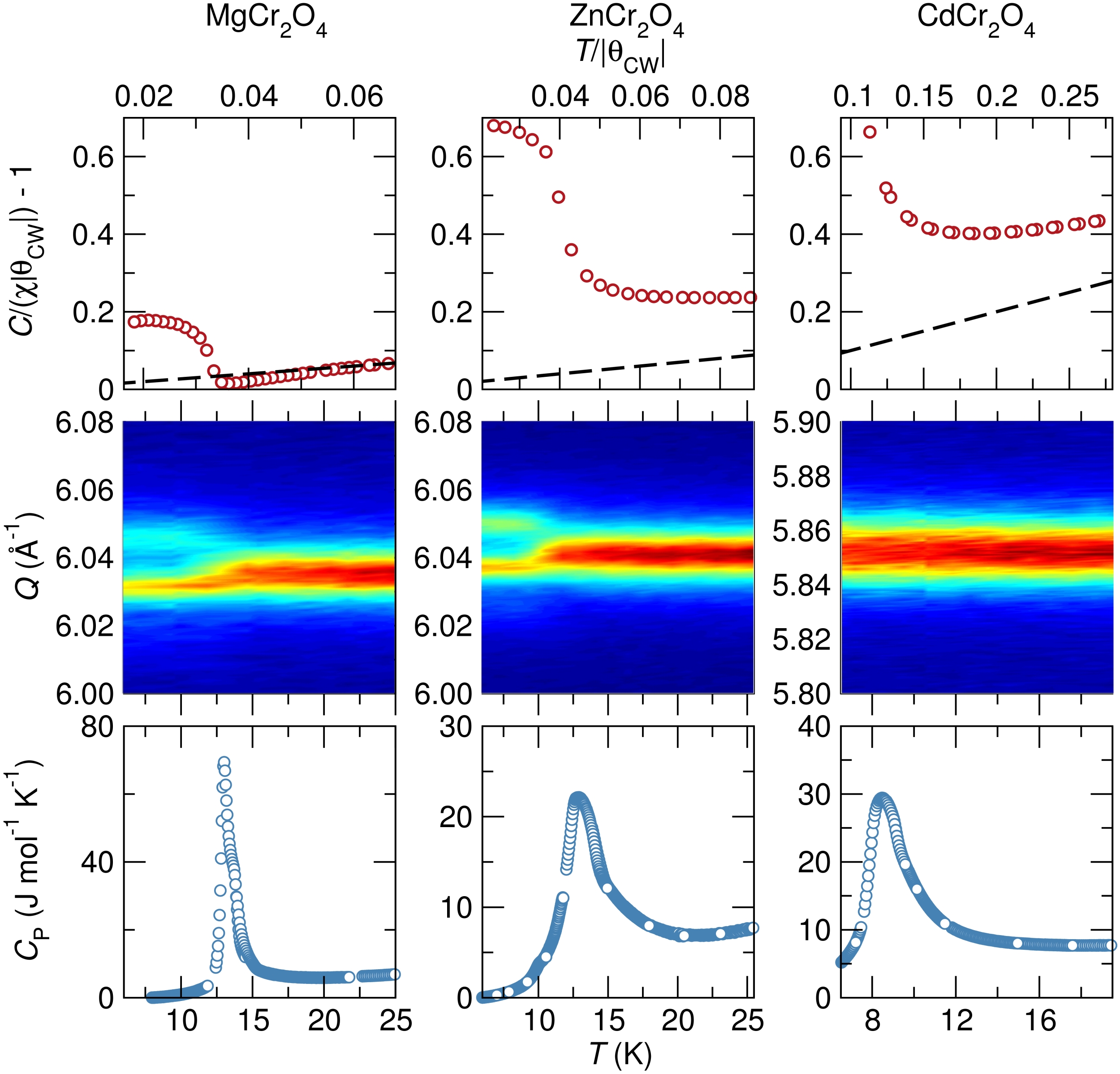}
\caption{\label{fig:aco} (Color online) Spin-Jahn-Teller distortions in $A$Cr$_2$O$_4$ 
spinels. The top panel shows the scaled inverse field-cooled susceptibility. The dashed black line models ideal paramagnetism. \mco\, and 
\zco\, were measured under a 1000 Oe field while \cco\, was measured in 6000 Oe. 
 Antiferromagnetic order is suppressed to low temperatures in \mco\,($T_N$\,=\,12.9\,K), 
\zco\,($T_N$\,=\,12.3\,K), and \cco\,($T_N$\,=\,7.86\,K). The splitting of high-symmetry
 cubic diffraction peaks into several low-symmetry peaks shows the onset of spin-driven 
structural distortions(middle panel). \cco\, shows a subtle structural distortion that 
is indicated by a slight decrease in intensity and increase in width of the high-symmetry
 peak. The bottom panel shows the change in entropy at the N\'eel temperature.}
\end{figure}

\mco\, was prepared by calcination of appropriate solution mixtures of the nitrates 
Mg(NO$_3$)$_2\cdot$6H$_2$O and Cr(NO$_3$)$_3\cdot$9H$_2$O at 1000$^{\circ}$\,C
for 10 hours. \zco\, and \cco\, were prepared by solid-state methods from ZnO, CdO, 
and Cr$_2$O$_3$ powders. Samples were annealed in the temperature range 800$^{\circ}$\,C to
1100$^{\circ}$\,C. A separate \zco\, sample was prepared in a Pt crucible by heating ZnO in 
an excess K$_2$Cr$_2$O$_7$ flux at 800$^{\circ}$\,C for 24 hours, 
followed by cooling at 15$^{\circ}$\,C/hr to room temperature. 
Samples were structurally characterized by 
high-resolution($\Delta Q/Q \leq 2 \times 10^{-4}$) synchrotron 
X-ray powder diffraction at temperatures between 6\,K and 295\,K. These measurements 
were performed at beamline 11-BM of the Advanced Photon Source, 
Argonne National Laboratory.
Structural models were refined against diffraction data
using the Rietveld method as implemented in the EXPGUI$/$GSAS software
program.\cite{toby_expgui_2001,larson_2000}  Atom positions for the low-symmetry
structures were obtained using the internet-server tool
ISODISPLACE.\cite{campbell_2006} Crystal distortions were analyzed using the
program VESTA.\cite{momma_vesta_2008} Magnetic properties were characterized
using a Quantum Design MPMS 5XL superconducting quantum interference device
(SQUID). Heat capacity measurements were performed using a Quantum Design
Physical Properties Measurement System. 

\begin{table}
\caption{\label{tab:magnetism} Magnetic parameters of \aco\, spinels }
\begin{ruledtabular}
\begin{tabular}{llllll}
&$T_N$ (K) &$\Theta_{CW}$ (K) & $f = \frac{|\Theta_{CW}|}{T_N}$ & $\mu_{exp}$ ($\mu_B$) & $\mu_{calc}$ ($\mu_B$)\\
\hline
\mco\, & 12.9 & $-$368 & 29 & 5.4 & 5.47\\
\zco\, & 12.3 & $-$288 & 23 & 5.2 & 5.47\\
\cco\, & 7.86 & $-$69.7 & 8.9 & 5.3 & 5.47\\
\end{tabular}
\end{ruledtabular}
\end{table}

At room temperature, the prepared \aco\, spinels are single phase homogeneous
compounds in the space group $Fd\overline{3}m$ with lattice parameters
8.33484(8)\,\AA\/ for \mco\, and 8.32765(8)\,\AA\/ for \zco\,;
\mco\, has a Cr$_2$O$_3$ impurity of 3.0 wt.\%. The \zco\, sample prepared from
K$_2$Cr$_2$O$_7$ flux has the cubic lattice constant  8.3288(2)\,\AA. The
cell parameters are in good agreement with previous reports.\cite{martin_2008} 
Scaled inverse field-cooled susceptibilities of \aco\, as described by the recast
Curie--Weiss equation are shown in the top panel of Fig.\,\ref{fig:aco}.\cite{melot_2009} 
Antiferromagnetic ordering occurs when $T/\Theta_{CW}$\,$\ll$\,1, indicating 
geometrically frustrated spin interactions (top axis scale of Fig.\,\ref{fig:aco}). 
Slight antiferromagnetic spin correlations are
observed above $T_N$ in \zco\, and \cco. We define $T_N$ as the temperature at which 
$d\chi_{ZFC}/dT$ is maximized. Magnetic
properties of \mco\, and \zco\, are extremely sensitive to non-stoichiometry.\cite{dutton_2011} The
magnetic properties of the samples presented here are tabulated in Table \ref{tab:magnetism}
and are in good agreement with earlier reports of highly stoichiometric 
compounds.\cite{dutton_2011,melot_2009}
Experimental magnetic moments of these compounds are within error of the
calculated effective moment of 5.47$\mu_B$ (Table \ref{tab:magnetism}). There is
a $\approx$0.3\,K thermal hysteresis between the zero-field-cooled and field-cooled
temperature dependent susceptibilities of the \aco\, spinels. We observe a
$\Theta_{CW}$ of -288\,K for \zco\, which is consistent with the earlier work by
Melot $et\,al.$\cite{melot_2009} but is lower than other $\Theta_{CW}$ values reported in the
literature.\cite{martin_2008,ramirez_1994}
The magnetic ordering transitions of \aco\, spinels are associated with changes in entropy (Figure
\ref{fig:aco}) and this agrees well with the earlier work of Klemme 
\textit{et al.}\cite{klemme_2000,klemme_2004} \zco\, and \cco\, have smooth
heat capacity anomalies while \mco\, has a sharp anomaly with a
shoulder feature that could indicate that its structural and magnetic
transitions occur at slightly different temperatures. 

 \begin{table}
\caption{\label{tab:table1} The low-temperature
structures of \mco\, and \zco\, as determined from Rietveld refinement of
high-resolution synchrotron X-ray powder diffraction data. All
atomic parameters were allowed to vary during the structural refinement except
for isotropic thermal parameters that are constrained to be the same for both
low-temperature phases.}

\centering
\begin{ruledtabular}
\begin{tabular}{llll}
 &MgCr$_2$O$_4$&ZnCr$_2$O$_4$&ZnCr$_2$O$_4$\footnote{\zco\, prepared from a flux of K$_2$Cr$_2$O$_7$}\\
\hline
$T$ (K) & 5.7 & 5.4 & 6.9 \\
$\lambda$ (\AA) & 0.413393 & 0.413399 & 0.413331 \\
space group & $I4_1/amd$& $I4_1/amd$& $I4_1/amd$\\
$Z$ & 4 &4&4\\
$a$ (\AA)& 5.89351(2) & 5.88753(1) & 5.88919(2) \\
$c$ (\AA)& 8.31503(7) & 8.30895(4) & 8.31703(5) \\
Vol (\AA$^3$) & 288.809(2) & 288.013(2)& 288.456(2)  \\
Mg/Zn & (0,$\frac{3}{4}$,$\frac{1}{8}$)& (0,$\frac{3}{4}$,$\frac{1}{8}$) & (0,$\frac{3}{4}$,$\frac{1}{8}$) \\
Cr & (0,0,$\frac{1}{2}$) &(0,0,$\frac{1}{2}$) & (0,0,$\frac{1}{2}$) \\
O & 0 & 0 &  0\\
 & 0.5240(3) &0.5250(4) &  0.5196(5)\\
 & 0.7391(2) &0.7379(4) &  0.7387(5)\\ 
wt. frac. & 0.42(0)& 0.43(0) &0.39(0)\\
\footnote{\label{note3}Determined from Scherrer analysis of well resolved peaks. This is the lower limit of crystallite size and assumes that all peak broadening is due to crystallite size}coherence length(nm) & 69.9 & 74.5& \\
space group & $Fddd$& $Fddd$& $Fddd$\\
$Z$ & 8 &8&8\\
$a$ (\AA)& 8.3041(2) & 8.3012(1) & 8.3059(9)\\
$b$ (\AA)& 8.3228(2) & 8.3144(1) & 8.3247(8)\\
$c$ (\AA)& 8.3526(2) & 8.3430(1) & 8.3415(0)\\
Vol (\AA$^3$) & 577.279(6) & 575.830(5)& 576.758(4) \\
Mg/Zn & ($\frac{1}{8}$,$\frac{1}{8}$,$\frac{1}{8}$)&  ($\frac{1}{8}$,$\frac{1}{8}$,$\frac{1}{8}$) & ($\frac{1}{8}$,$\frac{1}{8}$,$\frac{1}{8}$) \\
Cr &($\frac{1}{2}$,$\frac{1}{2}$,$\frac{1}{2}$)& ($\frac{1}{2}$,$\frac{1}{2}$,$\frac{1}{2}$) & ($\frac{1}{2}$,$\frac{1}{2}$,$\frac{1}{2}$)  \\
O & 0.26130(4) &0.26466(4) & 0.26092(3) \\
 & 0.26135(4) &0.25722(7) & 0.26639(5) \\
 & 0.26093(2) &0.26193(5) & 0.26015(3) \\
wt. frac. & 0.55(0)& 0.57(0)&0.61(0)\\
$^{b}$coherence length(nm) & 40.4 & 35.6& \\
Mg/Zn $U_{iso}$& 0.00290(1)&  0.00254(5) & 0.00305(4) \\
Cr $U_{iso}$& 0.00167(4) & 0.00074(5) & 0.00164(4)\\
O $U_{iso}$ & 0.00115(1)  & 0.00458(2) &  0.00321(2)\\ 
$\chi^2$  & 2.903 & 3.673 &  1.709\\
$R_{wp}$& 0.0331 & 0.0582 & 0.0823\\ 
\end{tabular}
\end{ruledtabular}
\end{table}
\begin{figure}

\includegraphics[scale=0.9]{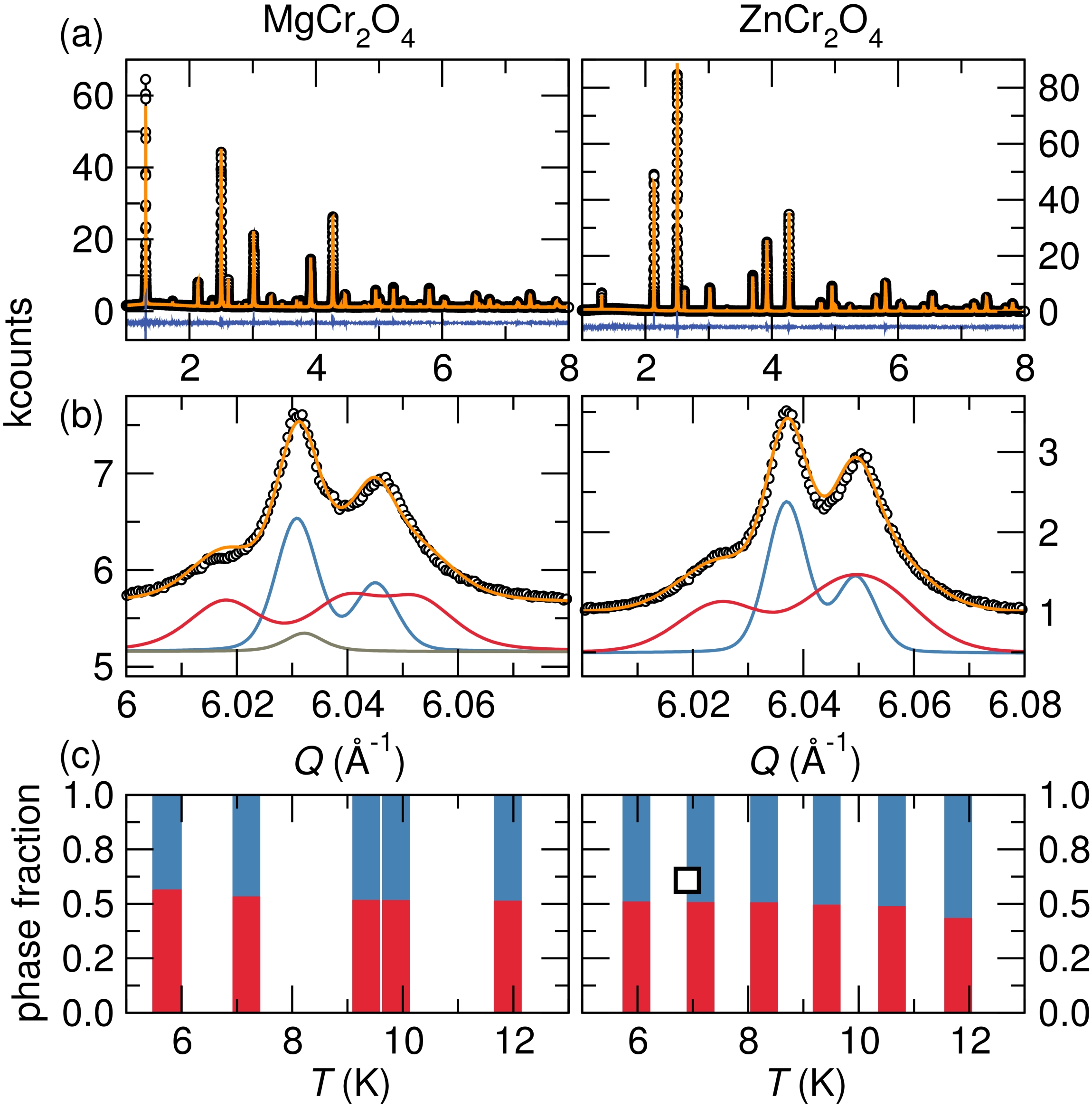}
\caption{\label{fig:lowsym} (Color online) Low-temperature diffraction and Rietveld 
refinement of \mco\,(left panel) and
\zco\,(right panel). (a) High-resolution synchrotron X-ray powder patterns
 collected at $\approx$6\,K and indexed to a two-phase model of tetragonal
$I4_1/amd$ and orthorhombic $Fddd$ symmetries [Data (black), combined
$I4_1/amd$ and $Fddd$ fit (orange), difference (blue)]. (b) The high-symmetry (800) peak 
splits into several $I4_1/amd$ and $Fddd$ reflections. The $I4_1/amd$ and $Fddd$ fits have been
offset from the data for clarity. [$I4_1/amd$ (blue), $Fddd$ (red), and
Cr$_2$O$_3$ impurity (grey)] (c) Nearly equal amounts of
$I4_1/amd$ (blue) and $Fddd$ (red) phases coexist below $T_N$; the $Fddd$ phase fraction 
increases slightly with decreasing $T$. The sample of
\zco\, prepared using K$_2$Cr$_2$O$_7$ flux shows a similar low-temperature
structure and its phase fractions are indicated by the square.}
\end{figure}

The cubic $Fd\overline{3}m$ (800)
diffraction peak of \mco\, and \zco\, splits into several low-symmetry peaks (Figure \ref{fig:aco}).
 \cco\/ on the other had, while displaying some peak broadening, remains well modelled by 
the high-temperature $Fd\overline{3}m$ space group even at 6.9\,K (Figure \ref{fig:aco}). 
Rietveld fits to the low-temperature synchrotron X-ray powder diffraction data of \mco\, and
\zco\, using structural models reported in the
literature\cite{ehrenberg_2002,martin_2008,lee_2007,kagomiya_2002} resulted in
regions of poorly fit intensity.  Similarly, the low-symmetry structures $F222$,
$C2/c$, and $I2/a$ could not model the data well. Group-subgroup
relations of the space group $Fd\overline{3}m$ yield the lower-symmetry groups
$I4_1/amd$ and $Fddd$. Individually, neither of these structural models can
reproduce the intensities and peak splittings observed in our low-temperature
diffraction patterns of \mco\, and \zco. However, we find that the diffraction data can
be well described by a two-phase model combining both tetragonal $I4_1/amd$ and
orthorhombic $Fddd$ structures [Fig.\,\ref{fig:lowsym}(a)]. This
refinement yields chemically reasonable and stable isotropic thermal
displacement parameters for both phases (Table \ref{tab:table1}).  
In Fig.\,\ref{fig:lowsym}(b), the low-temperature peak splitting of the cubic
$Fd\bar{3}m$ (800) reflection is deconvoluted into contributions from the
$I4_1/amd$ and $Fddd$ phases. 

Nearly equal fractions of the two phases coexist
in the low-temperature nuclear structures of \mco\, and \zco\, [Figure
\ref{fig:lowsym}(c)]. Employing the Thompson-Cox-Hastings pseudo-voigt profile function, we observe a slight increase
of the $Fddd$ phase fraction
with a decrease in temperature below $T_N$ for both \mco\, and \zco. 
While the estimated standard deviations suggest rather accurate
phase fractions(Table \ref{tab:table1}), separate refinements employing different profile functions show 
variations of up to 10$\%$. Scherrer analysis of deconvoluted $I4_1/amd$ and $Fddd$ peaks shown in Fig. \ref{fig:lowsym}(b) yield 
larger coherence lengths in the tetragonal phases of \mco\, and \zco\, compared with the orthorhombic phases(Table \ref{tab:table1}).
Williamson--Hall analysis yields room temperature 
crystallite sizes of 118\,nm in \mco\, and 200\,nm in \zco. The analysis reveals that larger $Fd\bar{3}m$ domains split into 
smaller domains of coexisting $I4_1/amd$ and $Fddd$ phases(Table \ref{tab:table1}).   
The two low-temperature phases coexist down to the lowest temperatures studied.

We have also examined a \zco\, sample prepared in a K$_2$Cr$_2$O$_7$ flux to
explore the effect of sample preparation conditions.  High-resolution
synchrotron X-ray diffraction measurements carried out at 7\,K reveal
that it is also described by a combination of both $I4_1/amd$ and $Fddd$. There
are subtle differences in the low-temperature phase composition of the
flux-prepared sample. Specifically, a slightly higher $Fddd$ phase fraction is
observed compared to the sample prepared by solid state methods[Figure
\ref{fig:lowsym}(c)].

\begin{figure}
\includegraphics[scale=0.8]{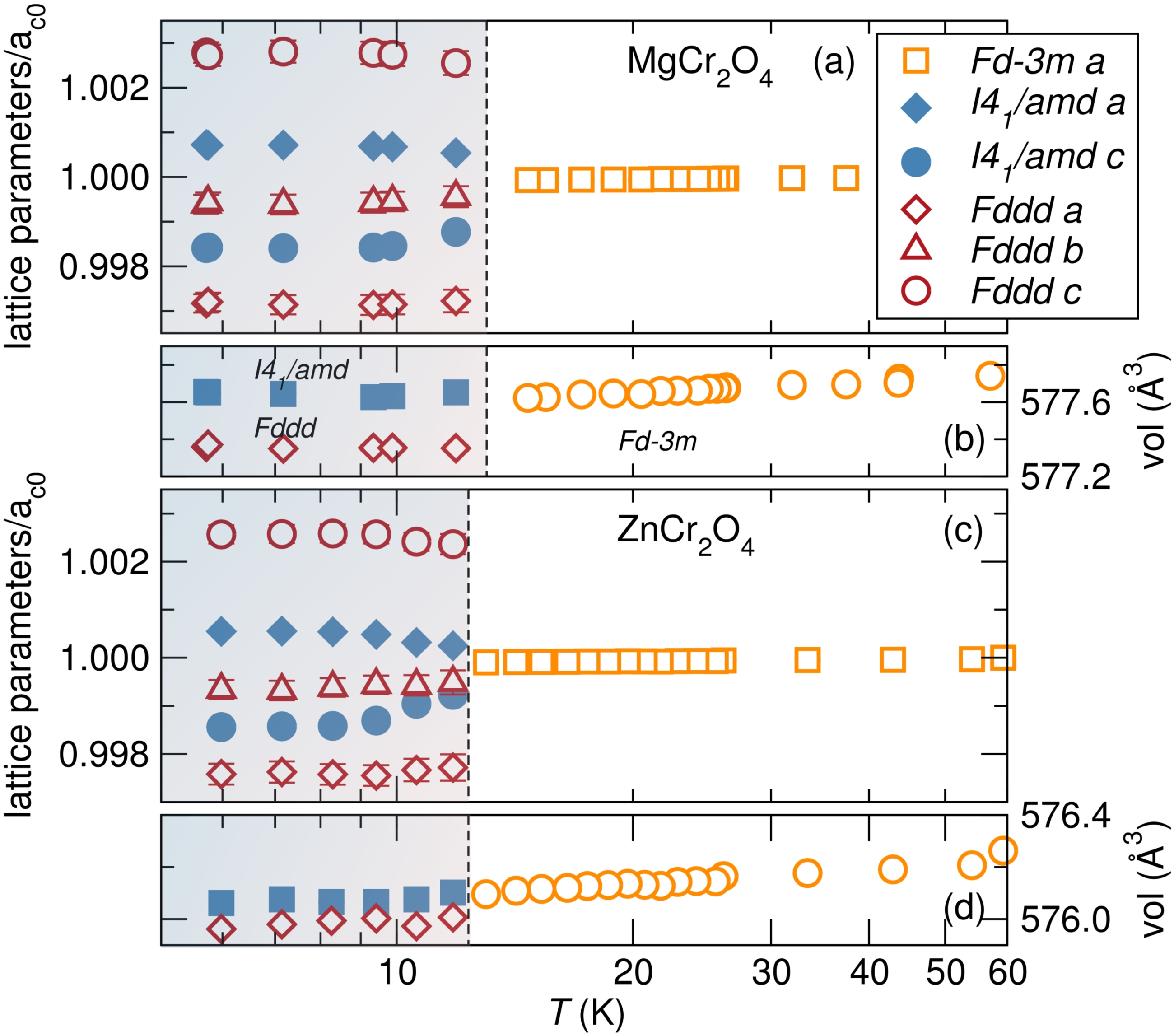}
\caption{\label{fig:MCOZCOlattice} (Color online) Temperature evolution of lattice
parameters in \mco\, and \zco\, through their magnetostructural distortions. 
The cubic lattice constants of
\mco\,(a) and \zco\,(c) separate into $I4_1/amd$ and $Fddd$ lattice constants at
$T_N$ = 12.9\,K and $T_N$ = 12.3\,K respectively. The lattice parameters of \mco\,
are normalized by the lattice constant at 57.1\,K ($a_{c0}$ = 8.32871\,\AA) while
the lattice constants of \zco\, are normalized by the lattice constant at
59.3\,K ($a_{c0}$ = 8.3216\,\AA). The $I4_1/amd$ lattice constants have been
multiplied by $\sqrt{2}$. A change in slope of the cell
volumes of \mco\,(b) and \zco\,(d) occurs at their respective $T_N$. 
In some cases, error bars are smaller than the symbols.}
\end{figure}

The $Fd\overline{3}m$ lattice parameter of \mco\, and \zco\, splits
abruptly into two $I4_1/amd$ and three $Fddd$ lattice constants at $T_N$, as
shown in Fig.\,\ref{fig:MCOZCOlattice}(a) and (c) respectively. The $Fddd$ $a$ and 
$c$ parameters of \mco\, and \zco\, show the
greatest distortion from cubic symmetry. The $Fddd$ phase of each compound has a smaller
volume than its $I4_1/amd$ counterpart, suggesting that $Fddd$ is
the lower energy structure. \mco\, and \zco\, undergo a first-order
structural transition at $T_N$ indicated by the change in slope of the cell 
volume [Figure \ref{fig:MCOZCOlattice}(b) and (d)], the onset
of a two-phase regime [Figure \ref{fig:lowsym}(b)], and the release of entropy(Figure \ref{fig:aco}). 
\begin{figure}
\includegraphics[scale=0.9]{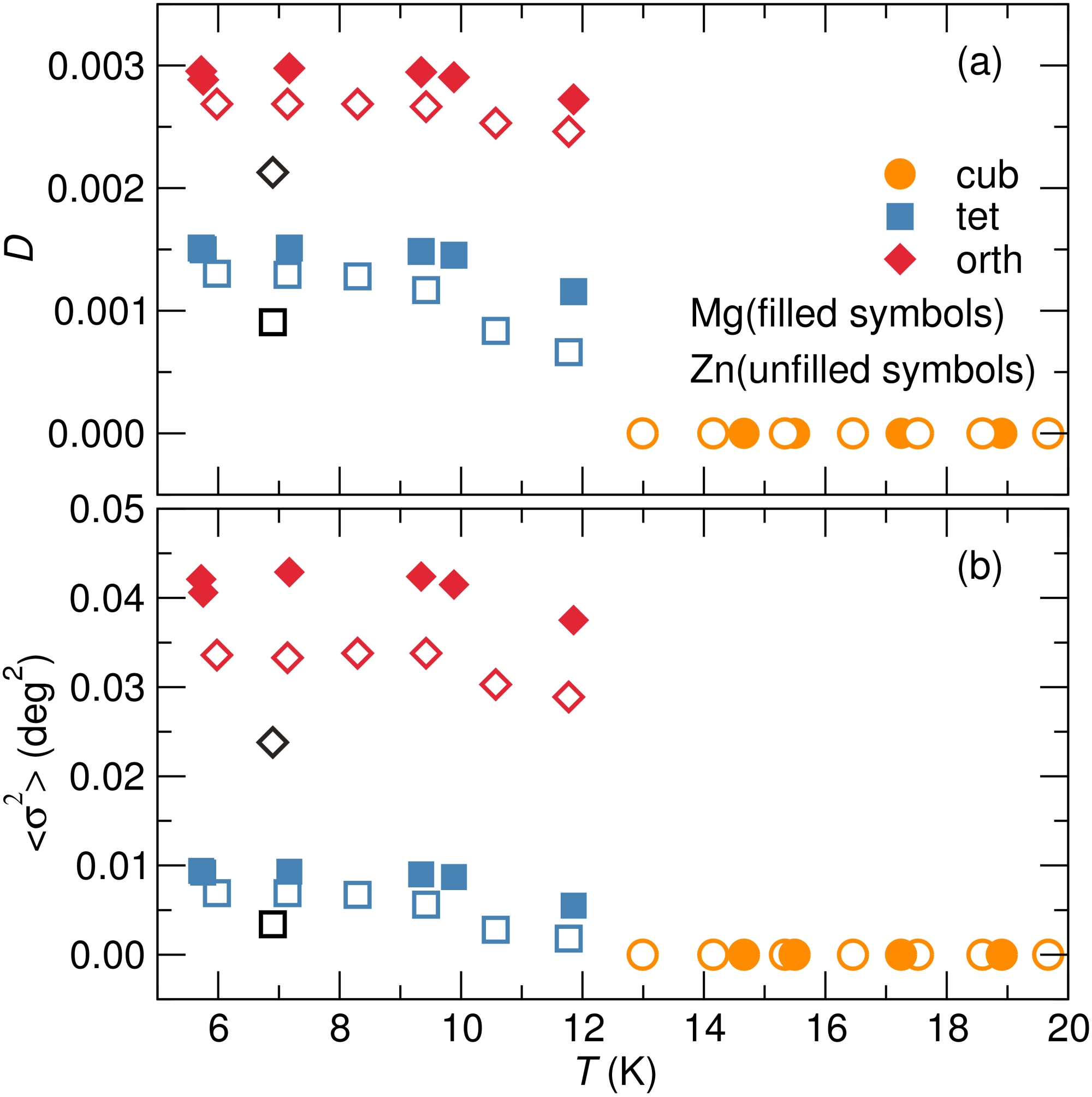}
\caption{\label{fig:distortion} (Color online) Distortion of Cr$_4$ tetrahedra
in the spin-Jahn-Teller phases of \mco\, and \zco. (a) At 12.9\,K and 12.3\,K, 
Cr-Cr bond distance distortions occur in
\mco\, and \zco\, respectively. (b) Distortion of Cr$_4$ angles occur below
$T_N$ in \mco\,  and \zco\,. The black symbols show distortions in the \zco\, sample
prepared from a K$_2$Cr$_2$O$_7$ flux.}
\end{figure}

The Cr$_4$ tetrahedra of \mco\, and \zco\, are distorted below $T_N$. We compute a
tetrahedral distortion index of Cr$_4$
tetrahedra, $D = 1/n \sum_{i=1}^n (l_i - \bar l)/\bar l$,  where $l_i$ is the $i$th Cr-Cr bond distance and $\bar l$ is the
average Cr-Cr bond distance.\cite{baur_1974} A larger $D$ is seen for the $Fddd$
phases of \mco\, and \zco\, in comparison to the 
$I4_1/amd$ phases [Figure\,\ref{fig:distortion}(a)]. Similarly, we compute an angle
variance of Cr$_4$ tetrahedra, $\sigma^2 = 1/(m-1)\sum_{i=1}^m
(\phi_i - \phi_0)^2$, where $\phi_0$ is the ideal tetrahedron angle of
109.47$^{\circ}$, $\phi_i$ is the measured angle, and $m$ is (the number of faces
of a tetrahedron)$\times$ 3/2.\cite{robinson_1971,momma_vesta_2008} A greater angle
variance occurs in the orthorhombic phases of \mco\, and \zco\, rather than in
the tetragonal phases. The \zco\, sample prepared from a K$_2$Cr$_2$O$_7$ flux has
less distortion of its Cr$_4$ tetrahedra compared with the solid state \zco\,
compound. Of the two compounds studied, the Cr$_4$ tetrahedra are more distorted
in \mco\, than in \zco. 

The spin-Jahn-Teller distortion of \mco\, and \zco\, resembles martensitic phase
transitions, which are displacive solid-solid transitions. These transformations
can be induced by varying temperature, involve changes in crystal symmetry
without a change in chemical composition, and show hysteresis. Volume changes
between the parent and product phases that occur at the spin-Jahn-Teller
distortion temperature could induce strains that result in a biphasic product.

It is important to consider whether a single low-symmetry space group could model the
data. Our refinements using $F222$, $C2/c$, or $I2/a$ were unable to
generate the observed peak separations. Analysis of the Cr$_4$ tetrahedra
distortions shows that the $Fddd$ phase is more distorted than the $I4_1/amd$
phase. Additionally, the $Fddd$ phase fraction increases slightly with a decrease
in temperature below $T_N$. The combination of these two effects would be
challenging to describe using a single low-symmetry structural model. Further, the
two-phase $I4_1/amd$ and $Fddd$ model is robust against changes in sample
preparation conditions. 

Phase coexistence following a phase transition is not unusual.
Compositional inhomogeneity contributes to multiple low-temperature phases in
Nd$_{0.5}$Sr$_{0.5}$MnO$_3$.\cite{woodward_1999} Similarly, complex phase behavior
featuring three coexisting phases occurs in the
relaxor-ferroelectric systems  Pb(Mg,Nb,Ti)O$_3$ due to internal strain, and
are proposed to be intrinsic to the system.\cite{noheda_2002} 
Distortion from $Fd\overline{3}m$ to $I4_1/amd$ symmetry, driven by orbital ordering, 
occurs in the related spinel compounds \nco\, and \cuco\,. This is followed by 
further distortion to $Fddd$ symmetry due to magnetostructural 
coupling.\cite{suchomel_2012}  The magnetostructural
distortions of \nco\, and \cuco\, are of the
same order of magnitude as the structural distortions we observe in \mco\, and
\zco. We can then consider that while tetragonal distortion alone may be sufficient to
lift spin degeneracy in \mco\, and \zco, magnetostructural coupling
could then drive further distortion from tetragonal to orthorhombic symmetry.
Although kinetics play a minor role in displacive transitions, the low temperatures of
spin-Jahn-Teller distortions in \mco\, and \zco\, could contribute to
stabilizing the metastable two-phase system reported here, preventing the emergence of a true
ground state.

In summary, we report coexisting $I4_1/amd$ and $Fddd$ phases 
in the spin-Jahn-Teller structures of \mco\, and \zco. 
Nearly equal phase fractions of the $I4_1/amd$ and the 
$Fddd$ phase coexist below $T_N$. The tetragonal phases have larger coherence lengths
than the orthorhombic phases.  Understanding material structure in these canonically 
frustrated systems has important consequences for unravelling their degenerate ground 
states wherein interesting physics is predicted and novel functional states may exist. 

We thank Jason Douglas for helpful discussions. 
MCK is supported by the Schlumberger Foundation Faculty for the Future
fellowship. PTB is supported by the National Science Foundation Graduate
Research Fellowship program. MWG is supported by a NSERC Postgraduate Scholarship and 
an International Fulbright Science \& Technology Award. RS, MCK, and PTB acknowledge 
the support of the NSF through the DMR 1105301. YIK is supported by National Research Foundation of
Korea (2012-0002868). We acknowledge the use of shared experimental facilities
of the Materials Research Laboratory: an NSF MRSEC, supported by NSF DMR
1121053. The 11-BM beamline at the Advanced Photon Source is supported by the
DOE, Office of Science, Office of Basic Energy Sciences, under Contract No.
DE-AC0206CH11357.

\bibliography{Kemei}

\begin{thebibliography}{27}
\expandafter\ifx\csname natexlab\endcsname\relax\def\natexlab#1{#1}\fi
\expandafter\ifx\csname bibnamefont\endcsname\relax
  \def\bibnamefont#1{#1}\fi
\expandafter\ifx\csname bibfnamefont\endcsname\relax
  \def\bibfnamefont#1{#1}\fi
\expandafter\ifx\csname citenamefont\endcsname\relax
  \def\citenamefont#1{#1}\fi
\expandafter\ifx\csname url\endcsname\relax
  \def\url#1{\texttt{#1}}\fi
\expandafter\ifx\csname urlprefix\endcsname\relax\def\urlprefix{URL }\fi
\providecommand{\bibinfo}[2]{#2}
\providecommand{\eprint}[2][]{\url{#2}}

\bibitem[{\citenamefont{Lee et~al.}(2000)\citenamefont{Lee, Broholm, Kim,
  Ratcliff, and Cheong}}]{lee_2000}
\bibinfo{author}{\bibfnamefont{S.-H.} \bibnamefont{Lee}},
  \bibinfo{author}{\bibfnamefont{C.}~\bibnamefont{Broholm}},
  \bibinfo{author}{\bibfnamefont{T.~H.} \bibnamefont{Kim}},
  \bibinfo{author}{\bibfnamefont{W.}~\bibnamefont{Ratcliff}}, \bibnamefont{and}
  \bibinfo{author}{\bibfnamefont{S.-W.} \bibnamefont{Cheong}},
  \bibinfo{journal}{Phys. Rev. Lett.} \textbf{\bibinfo{volume}{84}},
  \bibinfo{pages}{3718} (\bibinfo{year}{2000}).

\bibitem[{\citenamefont{Tchernyshyov et~al.}(2002)\citenamefont{Tchernyshyov,
  Moessner, and Sondhi}}]{tchernyshyov_2002}
\bibinfo{author}{\bibfnamefont{O.}~\bibnamefont{Tchernyshyov}},
  \bibinfo{author}{\bibfnamefont{R.}~\bibnamefont{Moessner}}, \bibnamefont{and}
  \bibinfo{author}{\bibfnamefont{S.~L.} \bibnamefont{Sondhi}},
  \bibinfo{journal}{Phys. Rev. Lett.} \textbf{\bibinfo{volume}{88}},
  \bibinfo{pages}{067203} (\bibinfo{year}{2002}).

\bibitem[{\citenamefont{Lee et~al.}(2007)\citenamefont{Lee, Gasparovic,
  Broholm, Matsuda, Chung, Kim, Ueda, Xu, Zschack, Kakurai et~al.}}]{lee_2007}
\bibinfo{author}{\bibfnamefont{S.-H.} \bibnamefont{Lee}},
  \bibinfo{author}{\bibfnamefont{G.}~\bibnamefont{Gasparovic}},
  \bibinfo{author}{\bibfnamefont{C.}~\bibnamefont{Broholm}},
  \bibinfo{author}{\bibfnamefont{M.}~\bibnamefont{Matsuda}},
  \bibinfo{author}{\bibfnamefont{J.-H.} \bibnamefont{Chung}},
  \bibinfo{author}{\bibfnamefont{Y.~J.} \bibnamefont{Kim}},
  \bibinfo{author}{\bibfnamefont{H.}~\bibnamefont{Ueda}},
  \bibinfo{author}{\bibfnamefont{G.}~\bibnamefont{Xu}},
  \bibinfo{author}{\bibfnamefont{P.}~\bibnamefont{Zschack}},
  \bibinfo{author}{\bibfnamefont{K.}~\bibnamefont{Kakurai}},
  \bibnamefont{et~al.}, \bibinfo{journal}{J. Phys.: Condens. Matter}
  \textbf{\bibinfo{volume}{19}}, \bibinfo{pages}{145259}
  (\bibinfo{year}{2007}).

\bibitem[{\citenamefont{Kagomiya et~al.}(2002)\citenamefont{Kagomiya, Sawa,
  Siratori, Kohn, Toki, Hata, and Kita}}]{kagomiya_2002}
\bibinfo{author}{\bibfnamefont{I.}~\bibnamefont{Kagomiya}},
  \bibinfo{author}{\bibfnamefont{H.}~\bibnamefont{Sawa}},
  \bibinfo{author}{\bibfnamefont{K.}~\bibnamefont{Siratori}},
  \bibinfo{author}{\bibfnamefont{K.}~\bibnamefont{Kohn}},
  \bibinfo{author}{\bibfnamefont{M.}~\bibnamefont{Toki}},
  \bibinfo{author}{\bibfnamefont{Y.}~\bibnamefont{Hata}}, \bibnamefont{and}
  \bibinfo{author}{\bibfnamefont{E.}~\bibnamefont{Kita}},
  \bibinfo{journal}{Ferroelectrics} \textbf{\bibinfo{volume}{268}},
  \bibinfo{pages}{327} (\bibinfo{year}{2002}).

\bibitem[{\citenamefont{Dunitz and Orgel}(1957)}]{dunitz_1957}
\bibinfo{author}{\bibfnamefont{J.~D.} \bibnamefont{Dunitz}} \bibnamefont{and}
  \bibinfo{author}{\bibfnamefont{L.~E.} \bibnamefont{Orgel}},
  \bibinfo{journal}{J. Phys. Chem. Solids} \textbf{\bibinfo{volume}{3}},
  \bibinfo{pages}{20} (\bibinfo{year}{1957}).

\bibitem[{\citenamefont{Rudolf et~al.}(2007)\citenamefont{Rudolf, Kant, Mayr,
  Hemberger, Tsurkan, and Loidl}}]{rudolf_2007}
\bibinfo{author}{\bibfnamefont{T.}~\bibnamefont{Rudolf}},
  \bibinfo{author}{\bibfnamefont{C.}~\bibnamefont{Kant}},
  \bibinfo{author}{\bibfnamefont{F.}~\bibnamefont{Mayr}},
  \bibinfo{author}{\bibfnamefont{J.}~\bibnamefont{Hemberger}},
  \bibinfo{author}{\bibfnamefont{V.}~\bibnamefont{Tsurkan}}, \bibnamefont{and}
  \bibinfo{author}{\bibfnamefont{A.}~\bibnamefont{Loidl}},
  \bibinfo{journal}{New J. Phys.} \textbf{\bibinfo{volume}{9}},
  \bibinfo{pages}{76} (\bibinfo{year}{2007}).

\bibitem[{\citenamefont{Ramirez}(1994)}]{ramirez_1994}
\bibinfo{author}{\bibfnamefont{A.~P.} \bibnamefont{Ramirez}},
  \bibinfo{journal}{Annu. Rev. Mater. Sci.} \textbf{\bibinfo{volume}{24}},
  \bibinfo{pages}{453} (\bibinfo{year}{1994}).

\bibitem[{\citenamefont{Ueda et~al.}(2006)\citenamefont{Ueda, Mitamura, Goto,
  and Ueda}}]{ueda_2006}
\bibinfo{author}{\bibfnamefont{H.}~\bibnamefont{Ueda}},
  \bibinfo{author}{\bibfnamefont{H.}~\bibnamefont{Mitamura}},
  \bibinfo{author}{\bibfnamefont{T.}~\bibnamefont{Goto}}, \bibnamefont{and}
  \bibinfo{author}{\bibfnamefont{Y.}~\bibnamefont{Ueda}},
  \bibinfo{journal}{Phys. Rev. B} \textbf{\bibinfo{volume}{73}},
  \bibinfo{pages}{094415} (\bibinfo{year}{2006}).

\bibitem[{\citenamefont{Martin et~al.}(2008)\citenamefont{Martin, Williams,
  Gordon, Klemme, and Attfield}}]{martin_2008}
\bibinfo{author}{\bibfnamefont{L.-S.} \bibnamefont{Martin}},
  \bibinfo{author}{\bibfnamefont{A.~J.} \bibnamefont{Williams}},
  \bibinfo{author}{\bibfnamefont{C.~D.} \bibnamefont{Gordon}},
  \bibinfo{author}{\bibfnamefont{S.}~\bibnamefont{Klemme}}, \bibnamefont{and}
  \bibinfo{author}{\bibfnamefont{J.~P.} \bibnamefont{Attfield}},
  \bibinfo{journal}{J. Phys.: Condens. Matter} \textbf{\bibinfo{volume}{20}},
  \bibinfo{pages}{104238} (\bibinfo{year}{2008}).

\bibitem[{\citenamefont{Rovers et~al.}(2002)\citenamefont{Rovers, Kyriakou,
  Dabkowska, Luke, Larkin, and Savici}}]{rovers_2002}
\bibinfo{author}{\bibfnamefont{M.~T.} \bibnamefont{Rovers}},
  \bibinfo{author}{\bibfnamefont{P.~P.} \bibnamefont{Kyriakou}},
  \bibinfo{author}{\bibfnamefont{H.~A.} \bibnamefont{Dabkowska}},
  \bibinfo{author}{\bibfnamefont{G.~M.} \bibnamefont{Luke}},
  \bibinfo{author}{\bibfnamefont{M.~I.} \bibnamefont{Larkin}},
  \bibnamefont{and} \bibinfo{author}{\bibfnamefont{A.~T.}
  \bibnamefont{Savici}}, \bibinfo{journal}{Phys. Rev. B}
  \textbf{\bibinfo{volume}{66}}, \bibinfo{pages}{174434}
  (\bibinfo{year}{2002}).

\bibitem[{\citenamefont{Chung et~al.}(2005)\citenamefont{Chung, Matsuda, Lee,
  Kakurai, Ueda, Sato, Takagi, Hong, and Park}}]{chung_2005}
\bibinfo{author}{\bibfnamefont{J.-H.} \bibnamefont{Chung}},
  \bibinfo{author}{\bibfnamefont{M.}~\bibnamefont{Matsuda}},
  \bibinfo{author}{\bibfnamefont{S.-H.} \bibnamefont{Lee}},
  \bibinfo{author}{\bibfnamefont{K.}~\bibnamefont{Kakurai}},
  \bibinfo{author}{\bibfnamefont{H.}~\bibnamefont{Ueda}},
  \bibinfo{author}{\bibfnamefont{T.~J.} \bibnamefont{Sato}},
  \bibinfo{author}{\bibfnamefont{H.}~\bibnamefont{Takagi}},
  \bibinfo{author}{\bibfnamefont{K.~P.} \bibnamefont{Hong}}, \bibnamefont{and}
  \bibinfo{author}{\bibfnamefont{S.}~\bibnamefont{Park}},
  \bibinfo{journal}{Phys. Rev. Lett.} \textbf{\bibinfo{volume}{95}},
  \bibinfo{pages}{247204} (\bibinfo{year}{2005}).

\bibitem[{\citenamefont{ValdesAguilar et~al.}(2008)\citenamefont{ValdesAguilar,
  Sushkov, Choi, Cheong, and Drew}}]{aguilar_2008}
\bibinfo{author}{\bibfnamefont{R.}~\bibnamefont{ValdesAguilar}},
  \bibinfo{author}{\bibfnamefont{A.~B.} \bibnamefont{Sushkov}},
  \bibinfo{author}{\bibfnamefont{Y.~J.} \bibnamefont{Choi}},
  \bibinfo{author}{\bibfnamefont{S.~W.} \bibnamefont{Cheong}},
  \bibnamefont{and} \bibinfo{author}{\bibfnamefont{H.~D.} \bibnamefont{Drew}},
  \bibinfo{journal}{Phys. Rev. B} \textbf{\bibinfo{volume}{77}},
  \bibinfo{pages}{092412} (\bibinfo{year}{2008}).

\bibitem[{\citenamefont{Ehrenberg et~al.}(2002)\citenamefont{Ehrenberg, Knapp,
  Baehtz, and Klemme}}]{ehrenberg_2002}
\bibinfo{author}{\bibfnamefont{H.}~\bibnamefont{Ehrenberg}},
  \bibinfo{author}{\bibfnamefont{M.}~\bibnamefont{Knapp}},
  \bibinfo{author}{\bibfnamefont{C.}~\bibnamefont{Baehtz}}, \bibnamefont{and}
  \bibinfo{author}{\bibfnamefont{S.}~\bibnamefont{Klemme}},
  \bibinfo{journal}{Powder Diffr.} \textbf{\bibinfo{volume}{17}},
  \bibinfo{pages}{230} (\bibinfo{year}{2002}).

\bibitem[{\citenamefont{Glazkov et~al.}(2009)\citenamefont{Glazkov, Farutin,
  Tsurkan, von Nidda, and Loidl}}]{glazkov_2008}
\bibinfo{author}{\bibfnamefont{V.~N.} \bibnamefont{Glazkov}},
  \bibinfo{author}{\bibfnamefont{A.~M.} \bibnamefont{Farutin}},
  \bibinfo{author}{\bibfnamefont{V.}~\bibnamefont{Tsurkan}},
  \bibinfo{author}{\bibfnamefont{H.-A.~K.} \bibnamefont{von Nidda}},
  \bibnamefont{and} \bibinfo{author}{\bibfnamefont{A.}~\bibnamefont{Loidl}},
  \bibinfo{journal}{J. Phys. Conf. Ser.} \textbf{\bibinfo{volume}{145}},
  \bibinfo{pages}{012030} (\bibinfo{year}{2009}).

\bibitem[{\citenamefont{Toby}(2001)}]{toby_expgui_2001}
\bibinfo{author}{\bibfnamefont{B.~H.} \bibnamefont{Toby}}, \bibinfo{journal}{J.
  Appl. Crystallogr.} \textbf{\bibinfo{volume}{34}}, \bibinfo{pages}{210}
  (\bibinfo{year}{2001}).

\bibitem[{\citenamefont{Larson and Dreele}(2000)}]{larson_2000}
\bibinfo{author}{\bibfnamefont{A.~C.} \bibnamefont{Larson}} \bibnamefont{and}
  \bibinfo{author}{\bibfnamefont{R.~B.~V.} \bibnamefont{Dreele}},
  \bibinfo{journal}{Los Alamos National Laboratory Report} pp.
  \bibinfo{pages}{86--748} (\bibinfo{year}{2000}).

\bibitem[{\citenamefont{Campbell et~al.}(2006)\citenamefont{Campbell, Stokes,
  and Tanner}}]{campbell_2006}
\bibinfo{author}{\bibfnamefont{B.~J.} \bibnamefont{Campbell}},
  \bibinfo{author}{\bibfnamefont{H.~T.} \bibnamefont{Stokes}},
  \bibnamefont{and} \bibinfo{author}{\bibfnamefont{D.~E.}
  \bibnamefont{Tanner}}, \bibinfo{journal}{J. Appl. Crystallogr.}
  \textbf{\bibinfo{volume}{39}}, \bibinfo{pages}{607} (\bibinfo{year}{2006}).

\bibitem[{\citenamefont{Momma and Izumi}(2008)}]{momma_vesta_2008}
\bibinfo{author}{\bibfnamefont{K.}~\bibnamefont{Momma}} \bibnamefont{and}
  \bibinfo{author}{\bibfnamefont{F.}~\bibnamefont{Izumi}}, \bibinfo{journal}{J.
  Appl. Crystallogr.} \textbf{\bibinfo{volume}{41}}, \bibinfo{pages}{653}
  (\bibinfo{year}{2008}).

\bibitem[{\citenamefont{Melot et~al.}(2009)\citenamefont{Melot, Drewes,
  Seshadri, Stoudenmire, and Ramirez}}]{melot_2009}
\bibinfo{author}{\bibfnamefont{B.~C.} \bibnamefont{Melot}},
  \bibinfo{author}{\bibfnamefont{J.~E.} \bibnamefont{Drewes}},
  \bibinfo{author}{\bibfnamefont{R.}~\bibnamefont{Seshadri}},
  \bibinfo{author}{\bibfnamefont{E.~M.} \bibnamefont{Stoudenmire}},
  \bibnamefont{and} \bibinfo{author}{\bibfnamefont{A.~P.}
  \bibnamefont{Ramirez}}, \bibinfo{journal}{J. Phys.: Condens. Matter}
  \textbf{\bibinfo{volume}{21}}, \bibinfo{pages}{216007}
  (\bibinfo{year}{2009}).

\bibitem[{\citenamefont{Dutton et~al.}(2011)\citenamefont{Dutton, Huang,
  Tchernyshyov, Broholm, and Cava}}]{dutton_2011}
\bibinfo{author}{\bibfnamefont{S.~E.} \bibnamefont{Dutton}},
  \bibinfo{author}{\bibfnamefont{Q.}~\bibnamefont{Huang}},
  \bibinfo{author}{\bibfnamefont{O.}~\bibnamefont{Tchernyshyov}},
  \bibinfo{author}{\bibfnamefont{C.~L.} \bibnamefont{Broholm}},
  \bibnamefont{and} \bibinfo{author}{\bibfnamefont{R.~J.} \bibnamefont{Cava}},
  \bibinfo{journal}{Phys. Rev. B} \textbf{\bibinfo{volume}{83}},
  \bibinfo{pages}{064407} (\bibinfo{year}{2011}).

\bibitem[{\citenamefont{Klemme et~al.}(2000)\citenamefont{Klemme, O'Neill,
  Schnelle, and Gmelin}}]{klemme_2000}
\bibinfo{author}{\bibfnamefont{S.}~\bibnamefont{Klemme}},
  \bibinfo{author}{\bibfnamefont{H.}~\bibnamefont{O'Neill}},
  \bibinfo{author}{\bibfnamefont{W.}~\bibnamefont{Schnelle}}, \bibnamefont{and}
  \bibinfo{author}{\bibfnamefont{E.}~\bibnamefont{Gmelin}},
  \bibinfo{journal}{Am. Mineral.} \textbf{\bibinfo{volume}{85}},
  \bibinfo{pages}{1688} (\bibinfo{year}{2000}).

\bibitem[{\citenamefont{Klemme and Miltenburg}(2004)}]{klemme_2004}
\bibinfo{author}{\bibfnamefont{S.}~\bibnamefont{Klemme}} \bibnamefont{and}
  \bibinfo{author}{\bibfnamefont{J.~V.} \bibnamefont{Miltenburg}},
  \bibinfo{journal}{Mineral. Mag.} \textbf{\bibinfo{volume}{68}},
  \bibinfo{pages}{515} (\bibinfo{year}{2004}).

\bibitem[{\citenamefont{Baur}(1974)}]{baur_1974}
\bibinfo{author}{\bibfnamefont{W.~H.} \bibnamefont{Baur}},
  \bibinfo{journal}{Acta Crystallogr., Sect. B: Struct. Sci}
  \textbf{\bibinfo{volume}{30}}, \bibinfo{pages}{1195} (\bibinfo{year}{1974}).

\bibitem[{\citenamefont{Robinson et~al.}(1971)\citenamefont{Robinson, Gibbs,
  and Ribbe}}]{robinson_1971}
\bibinfo{author}{\bibfnamefont{K.}~\bibnamefont{Robinson}},
  \bibinfo{author}{\bibfnamefont{G.~V.} \bibnamefont{Gibbs}}, \bibnamefont{and}
  \bibinfo{author}{\bibfnamefont{P.~H.} \bibnamefont{Ribbe}},
  \bibinfo{journal}{Science} \textbf{\bibinfo{volume}{172}},
  \bibinfo{pages}{567} (\bibinfo{year}{1971}).

\bibitem[{\citenamefont{Woodward et~al.}(1999)\citenamefont{Woodward, Cox,
  Vogt, Rao, and Cheetham}}]{woodward_1999}
\bibinfo{author}{\bibfnamefont{P.~M.} \bibnamefont{Woodward}},
  \bibinfo{author}{\bibfnamefont{D.~E.} \bibnamefont{Cox}},
  \bibinfo{author}{\bibfnamefont{T.}~\bibnamefont{Vogt}},
  \bibinfo{author}{\bibfnamefont{C.~N.~R.} \bibnamefont{Rao}},
  \bibnamefont{and} \bibinfo{author}{\bibfnamefont{A.~K.}
  \bibnamefont{Cheetham}}, \bibinfo{journal}{Chem. Mater.}
  \textbf{\bibinfo{volume}{11}}, \bibinfo{pages}{3528} (\bibinfo{year}{1999}).

\bibitem[{\citenamefont{Noheda et~al.}(2002)\citenamefont{Noheda, Cox, Shirane,
  Gao, and Ye}}]{noheda_2002}
\bibinfo{author}{\bibfnamefont{B.}~\bibnamefont{Noheda}},
  \bibinfo{author}{\bibfnamefont{D.~E.} \bibnamefont{Cox}},
  \bibinfo{author}{\bibfnamefont{G.}~\bibnamefont{Shirane}},
  \bibinfo{author}{\bibfnamefont{J.}~\bibnamefont{Gao}}, \bibnamefont{and}
  \bibinfo{author}{\bibfnamefont{Z.-G.} \bibnamefont{Ye}},
  \bibinfo{journal}{Phys. Rev. B} \textbf{\bibinfo{volume}{66}},
  \bibinfo{pages}{054104} (\bibinfo{year}{2002}).

\bibitem[{\citenamefont{Suchomel et~al.}(2012)\citenamefont{Suchomel,
  Shoemaker, Ribaud, Kemei, and Seshadri}}]{suchomel_2012}
\bibinfo{author}{\bibfnamefont{M.~R.} \bibnamefont{Suchomel}},
  \bibinfo{author}{\bibfnamefont{D.~P.} \bibnamefont{Shoemaker}},
  \bibinfo{author}{\bibfnamefont{L.}~\bibnamefont{Ribaud}},
  \bibinfo{author}{\bibfnamefont{M.~C.} \bibnamefont{Kemei}}, \bibnamefont{and}
  \bibinfo{author}{\bibfnamefont{R.}~\bibnamefont{Seshadri}},
  \bibinfo{journal}{Phys. Rev. B} \textbf{\bibinfo{volume}{86}},
  \bibinfo{pages}{054406} (\bibinfo{year}{2012}).

\end{thebibliography}
\end{document}